\documentclass[12pt]{article}
\usepackage{jeosman}
\usepackage{epsfig}
\usepackage{amsmath}
\usepackage{amssymb}

\begin{document}

\title{Highly dispersive photonic band-gap-edge optofluidic biosensors}

\author{Sanshui Xiao and Niels Asger Mortensen}

\address{MIC -- Department of Micro and Nanotechnology, NanoDTU,\\Technical
University of Denmark, DK-2800 Kongens Lyngby, Denmark}

\email{sanshui.xiao@mic.dtu.dk}


\begin{abstract}
Highly dispersive photonic band-gap-edge optofluidic biosensors are
studied theoretically. We demonstrate that these structures are
strongly sensitive to the refractive index of the liquid, which is
used to tune dispersion of the photonic crystal. The upper frequency
band-gap edge shifts about 1.8~nm for $\delta n=0.002$, which is
quite sensitive. Results from transmission spectra agree well with
those obtained from the band structure theory.
\end{abstract}

\maketitle
\section{Introduction}
Photonic crystals (PhCs) are attractive optical materials for
controlling and manipulating the flow of
light~\cite{John:1987,Yab:1987,Joannopoulos:1995}. One well-know
property is the existence of photonic band gaps, i.e., regions of
frequencies where electromagnetic wave propagation is prohibited.
Another equally important aspect of a PhC is the existence of an
unconventional ultra-strong dispersion. Such an ultra-strong
dispersion was firstly reported by Lin {\it et al.} and demonstrated
experimentally in the millimeter-wave spectrum~\cite{Lin:1996}.
Kosaka {\it et al.} subsequently demonstrated the superprism effect
in a highly dispersive photonic microstructure~\cite{Kosaka:1998}.
These unusual properties provide an exciting possibility for
obtaining microphotonic and nanophotonic devices that can focus,
disperse, switch, and steer light.

Optofluidics, the marriage of nano-photonics and micro-fluidics,
refers to a class of optical systems that integrate optical and
fluidic devices~\cite{Psaltis:2006}. Due to unique properties of
fluids, such integration provides a new way for dynamic manipulation
of optical properties and shows many potential
applications~\cite{Domachuk:2004,Erickson:2006,Galas:2005,Grillet:2004,Kurt:2005,GersborgHansen:2005,Li:2006,GersborgHansen:2006}.
In particular, PhCs are interesting for optofluidics since they
naturally have voids where fluids can be injected. Optical
properties of the PhC can be easily reconfigured by selectively
filling specific voids with liquid. Chow {\it et al.} demonstrated
an ultra compact biosensor employing a two-dimensional (2D) photonic
crystal microcavity~\cite{Chow:2004}. Recently, we proposed simple
biosensor structures based on highly dispersive PhC
waveguides~\cite{Xiao:2006}. In this paper we will propose biosensor
structures based on complete PhCs and the strong dispersion occuring
near the Brillouin zone. In particular, the bandgap edges of the
PhCs are strongly sensitive to the refractive index of the liquid
which is used to tune the dispersion of the PhC. The suggested
structures show a potential for biochemical sensing applications.

\section{Biosensor Structures and Results}

\begin{figure}[t!]
\begin{center}
\epsfig{file=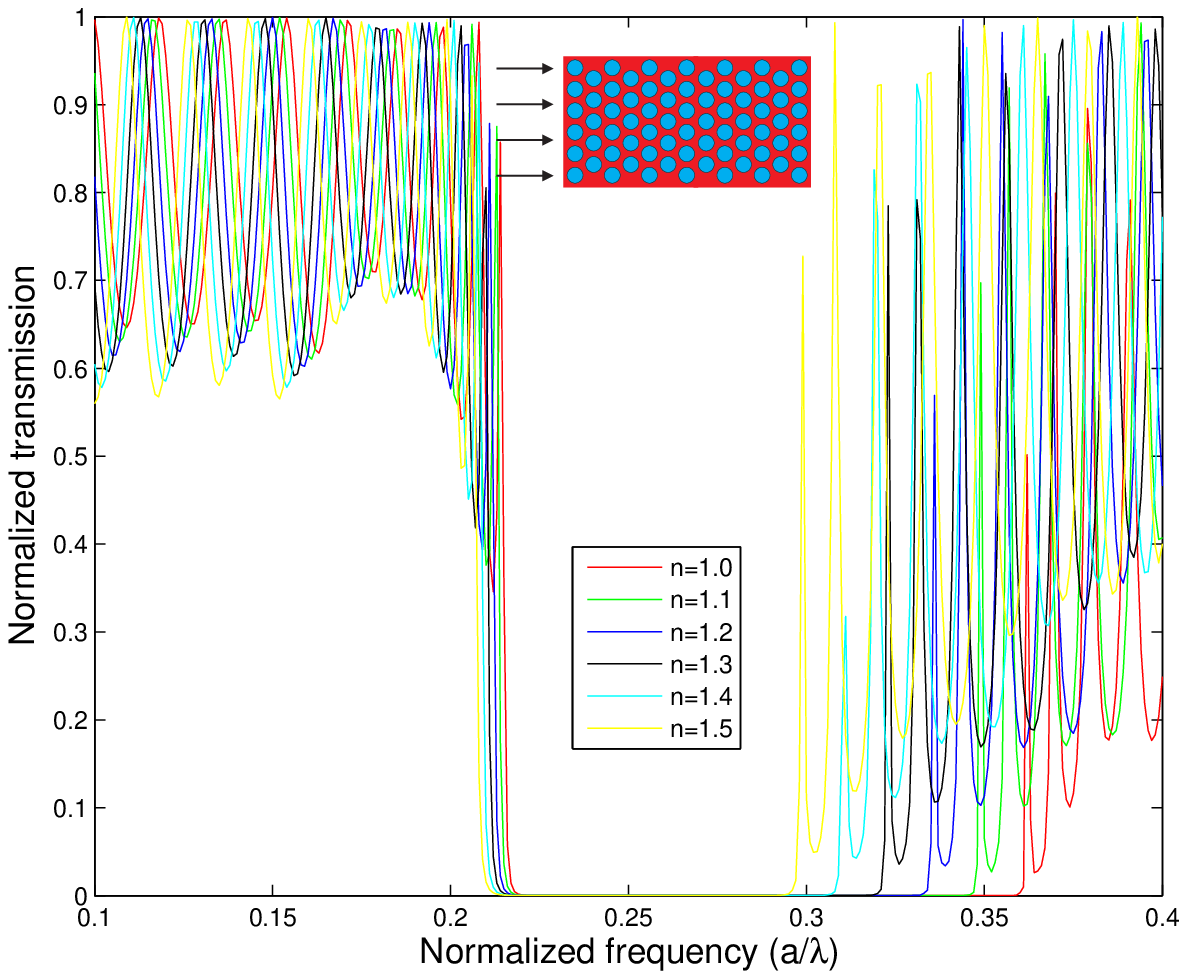,width=1 \columnwidth,clip,angle=0}
\end{center}
\caption{Transmission spectra for the light normally incident into
a triangular PhC, see inset, with air holes being filled by
different liquids with refractive indices varying from n=$1.0$ to
$1.5$ in steps of $0.1$. The PhC is a triangular lattice with
holes extending through a high-index $\varepsilon=10.5$ dielectric
slab and the radius of holes is $0.36a$, where $a$ is the lattice
constant.} \label{Trans-Tri}
\end{figure}

Let us first consider a 2D triangular PhC with air holes extending
through a high index $\varepsilon=10.5$ dielectric material, shown
in the inset of Fig.~\ref{Trans-Tri}. The holes have a radius of
$0.36a$, where $a$ is the lattice constant. Here, we will focus our
study on transmission spectra of the PhC with air holes being filled
with different liquids. It was shown in our previous work that both
the surface termination and surface direction of the PhC are
critical for high transmission (i.e., coupling) at an interface
between air and the PhC~\cite{Xiao:2004,Ruan:2005}. Consider the
TE-polarized (magnetic field parallel to the air holes) light
normally incident into the PhC. To enhance the coupling at the
interface, we choose to couple light to the PhC along the $\Gamma M$
direction, i.e., the surface of the PhC slab is along $\Gamma K$
direction. The symmetric PhC slab is composed of 11 layers along the
$\Gamma M$ direction and the distance (surface termination) between
right boundary and the centers of the first right holes is $0.5a$.
Transmission spectra for the PhC are obtained using the 2D
finite-difference time-domain (FDTD) method~\cite{TafloveFDTD}. For
this case, we use the periodic condition in $\Gamma K$ direction and
perfectly matched layers in the $\Gamma M$ direction
\cite{Berenger:1994} as the numerical boundary treatment.
Figure~\ref{Trans-Tri} shows transmission spectra for the PhC with
air holes being filled by different liquids with the refractive
index increasing from $n=1.0$ to $n=1.5$ in steps of $\delta n=0.1$.
One can see clearly that there exist band gaps for the PhCs.
Transmissions outside the band gaps are quite large and close to
unity for some frequencies. Peaks in the transmissions arise from
the Fabry--Perot oscillations from the two boundaries and the shifts
of peaks are due to the change of the effective index of the PhC
when filling air holes with different liquids.

\begin{figure}[t!]
\begin{center}
\epsfig{file=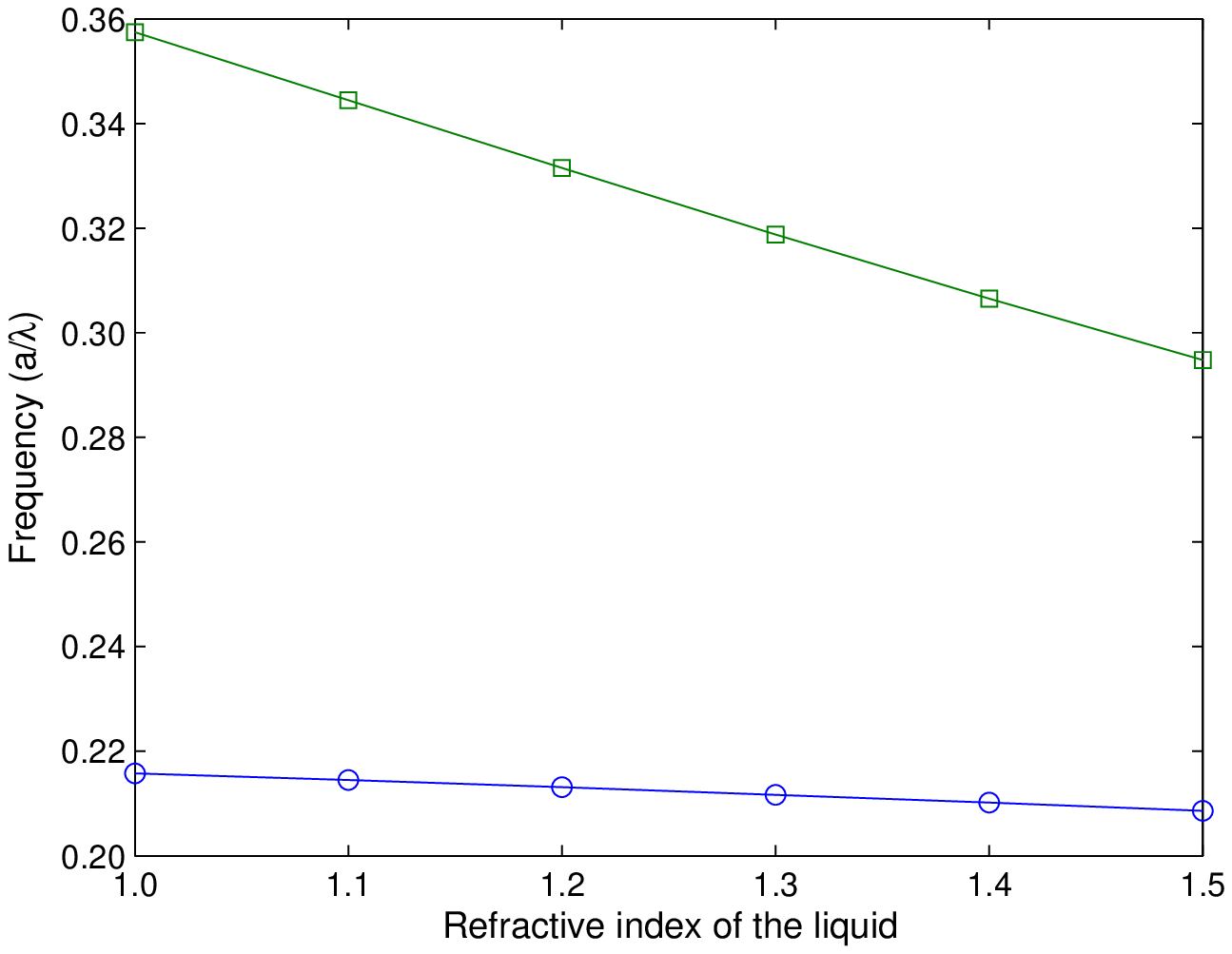,width=1\columnwidth,clip,angle=0}
\end{center}
\caption{Band-gap edges as a function of the refractive index for
the filled liquid. } \label{BandEdge-Tri}
\end{figure}

For the present application we are not interested in the details
of the Fabry--Perot pattern in Fig.~\ref{Trans-Tri}, but rather
the spectral position of the band-gap edge which is a highly
sensitive measure for changes in the refractive index of the
liquid. To see it more clearly, the change of the band-gap edge as
a function of the refractive index of the liquid is shown in
Fig.~\ref{BandEdge-Tri}. As seen, the low-frequency mode-gap edges
slightly change with the refractive index of the liquid. However,
the high-frequency mode-gap edge is strongly dependent on the
refractive index of the liquid, as shown by squares in
Fig.~\ref{BandEdge-Tri}. As an example, the high-frequency
band-gap edge shifts $\delta(a/\lambda)=0.012987$ when the air
holes (with index $n=1$) are filled by a liquid of index $n=1.1$.
For comparison, it is only $\delta(a/\lambda)=0.001278$ for the
low-frequency band-gap shift. Now, consider a commercial silicone
fluid with a calibrated refractive-index accuracy of $\delta
n=0.002$, as mentioned in Ref.~\cite{Chow:2004}, where the
refractive index of the liquid varies from $n=1.446$ to $1.454$ in
increments of $0.002$. For the working wavelength around
$1.55\,{\rm \mu m}$ (here we choose $a=450\,{\rm nm}$), the
high-frequency band-gap edge shifts up to 1.17~nm for $\delta
n=0.002$, while 0.33~nm for the low-frequency band-gap edge. For
comparison, we note that the shift in resonant wavelength for the
high-quality-factor PhC cavity is about $0.4$ nm for $\delta
n=0.002$~\cite{Chow:2004}. The above results demonstrate that even
such a simple PhC has potential applications as a sensitive
biosensor.

\begin{figure}[t!]
\begin{center}
\epsfig{file=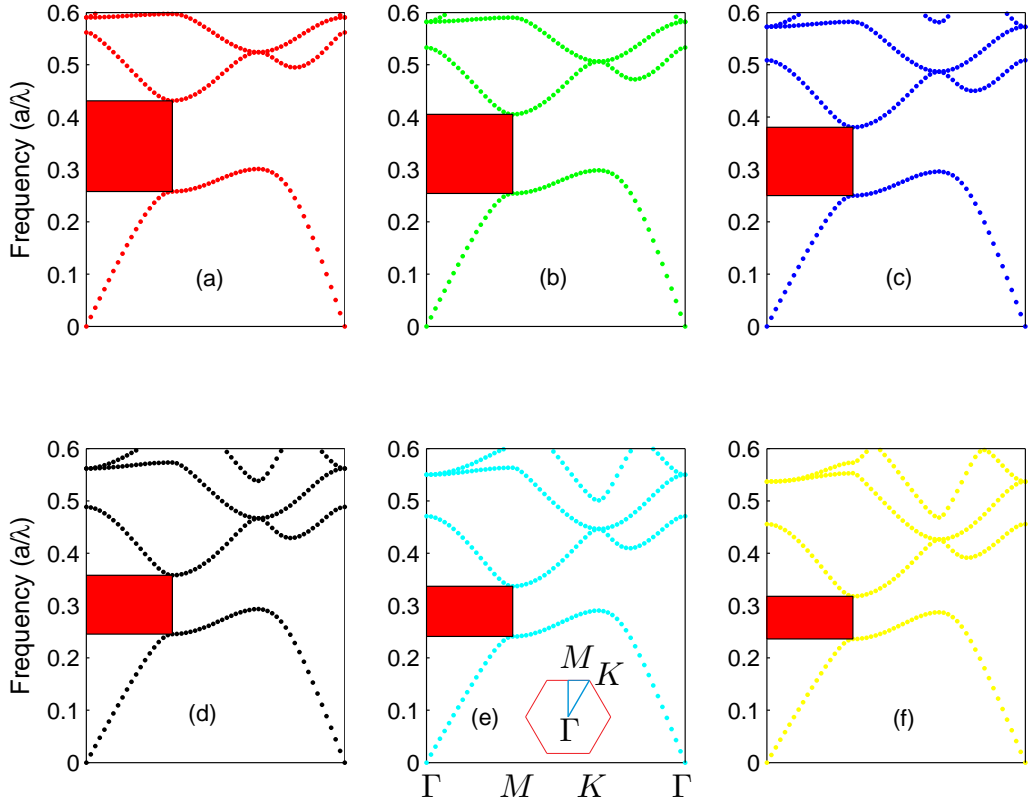,width=1\columnwidth,clip,angle=0}
\end{center}
\caption{Dispersion of the triangular photonic crystal shown in
the inset of Fig. 1(a), where the air holes are filled by liquids
with refractive indices varying from n=1.0 to 1.5 in steps of
$\delta n=0.1$. The red regions denote the bandgap regions along
$\Gamma M$ direction. } \label{Band-Tri}
\end{figure}

\begin{figure}[t!]
\begin{center}
\epsfig{file=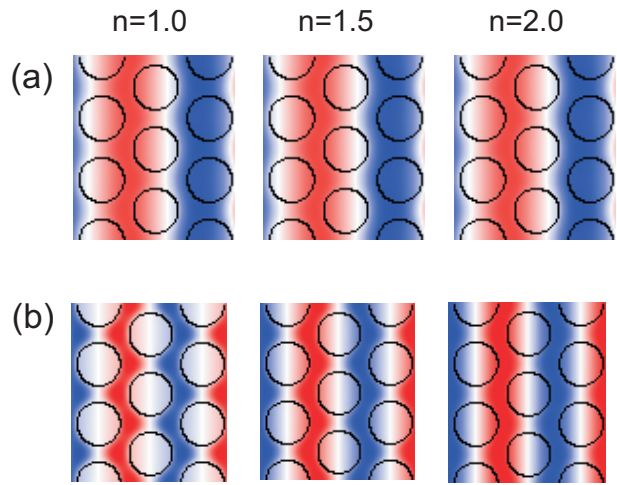,width=0.5\columnwidth,clip,angle=0}
\end{center}
\caption{Bloch mode at the band edge ($M$ point, $k=\pi/a$) for
the (a) first band (b) second band, in which air holes in the PhC
are filled by the liquid with a refractive index of $n=1.0$,
$1.5$,and $2.0$, respectively.} \label{Field}
\end{figure}

To further elucidate the physics behind the strong sensitivity, we
next support the picture by dispersion calculations for the PhC. For
this purpose we use a Block-iterative frequency-domain
method~\cite{Johnson:2001}. The dispersion of the PhC, in absence or
presence of a fluid, is shown in Fig.~\ref{Band-Tri}.
Figure~\ref{Band-Tri} (a)-(f) summarize the dispersions for the PhC,
where the air holes are filled by a liquid with a varying refractive
index. One can clearly see that these PhCs have photonic band gaps
for the TE polarization, which are related to the band gaps in
Fig.~\ref{Trans-Tri}, though the band-gap regions appear slightly
different with those obtained from the transmission spectra. Note
that the band gaps in Fig.~\ref{Trans-Tri} are larger than those in
Fig.~\ref{Band-Tri}. This is because the band gaps in
Fig.~\ref{Band-Tri} are for all incident directions while the band
gaps in Fig.~\ref{Trans-Tri} are only for propagation along the
normal direction. From Figs.~\ref{Trans-Tri} and~\ref{Band-Tri}, we
also observe that the position of the gap in the transmission
spectra, which are obtained for plane electromagnetic waves incident
normally on the PhC, agree very well with the position of the gaps
in the frequency band structure of the corresponding infinite
crystal along the $\Gamma M$ direction (denoted by red regions).
When increasing the refractive index of the liquid [going from panel
(a) toward panel (f)], the high-frequency band-gap edge is
significantly downward shifted, while the low-frequency band-gap
edge slightly decreases. We emphasize that all results obtained from
band structures are consistent with those from the transmission
spectra. The sensitivity of this structure is mainly attributed to
the strong dispersion of the PhC mode. Figures~\ref{Field}(a) and
~\ref{Field}(b) show the first and second PhC Bloch modes at the
band-gap edge ($M$ point, $k=\pi/a$), where air holes in the PhC are
filled by the liquid with a refractive index of $n=1.0$, $1.5$,and
$2.0$, respectively. As seen, the low-frequency band-edge Bloch mode
hardly changes with varying refractive index of the liquid. However,
for the high-frequency band-edge mode, the ratio of the energy in
holes becomes lager as the refractive index of the liquid increases,
i.e., this mode distribution is strongly dependent on the refractive
index of the liquid, which is in agreement with the results in
Figs.~\ref{Trans-Tri}-\ref{Band-Tri}. A somewhat similar structure
has been realized experimentally by Okamoto {\it et
al.}~\cite{Okamoto:2005}, but in this paper we have studied the
shift for the band-gap edges in details an offered a physical
explanation for the shifts. Besides, compared to the device in
Ref.~\cite{Okamoto:2005}, our proposed device offers a better
resolution.

\begin{figure}[t!]
\begin{center}
\epsfig{file=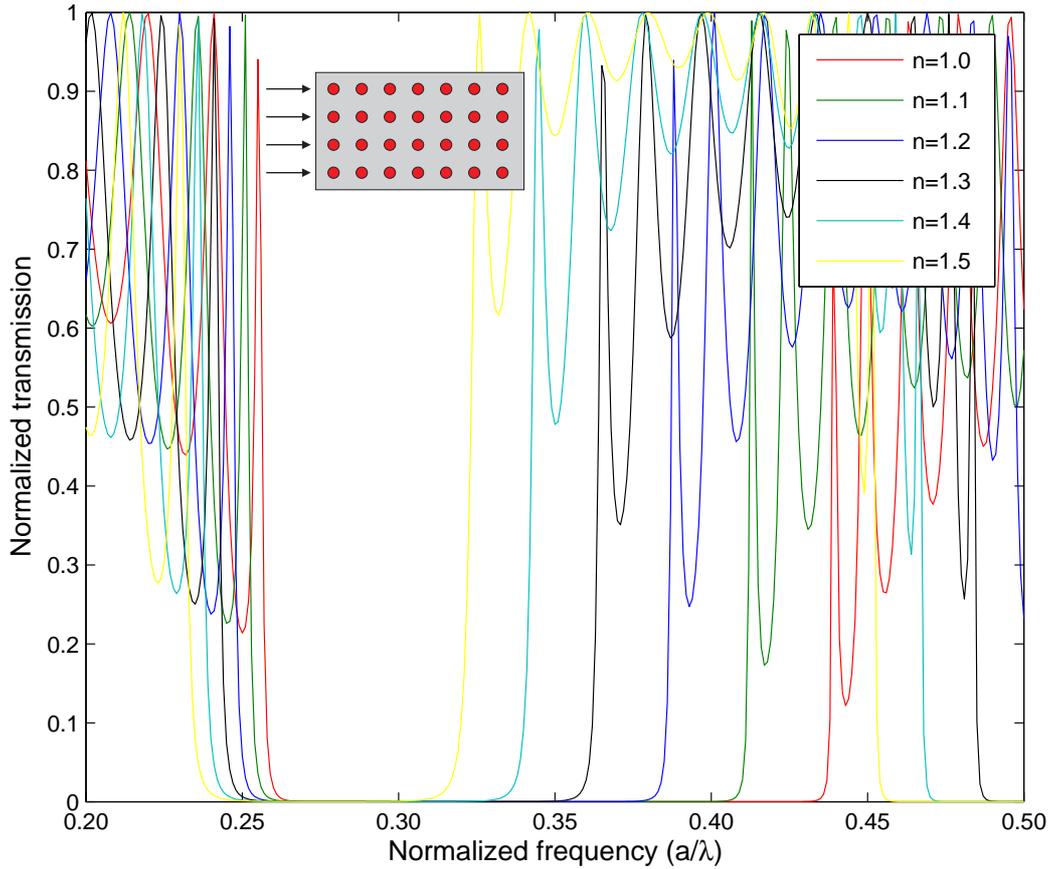,width=1 \columnwidth,clip,angle=0}
\end{center}
\caption{Transmission spectra for the light normally incident into
a square PhC, see inset, with the background being filled by
different liquids with refractive indices varying from n=$1.0$ to
$1.5$ in steps of $0.1$. The PhC is a square lattice with
dielectric rods in air. The rods have a refractive index of
$\varepsilon=10.5$ and the radius of $0.2a$.} \label{Trans-Squ}
\end{figure}

\begin{figure}[t!]
\begin{center}
\epsfig{file=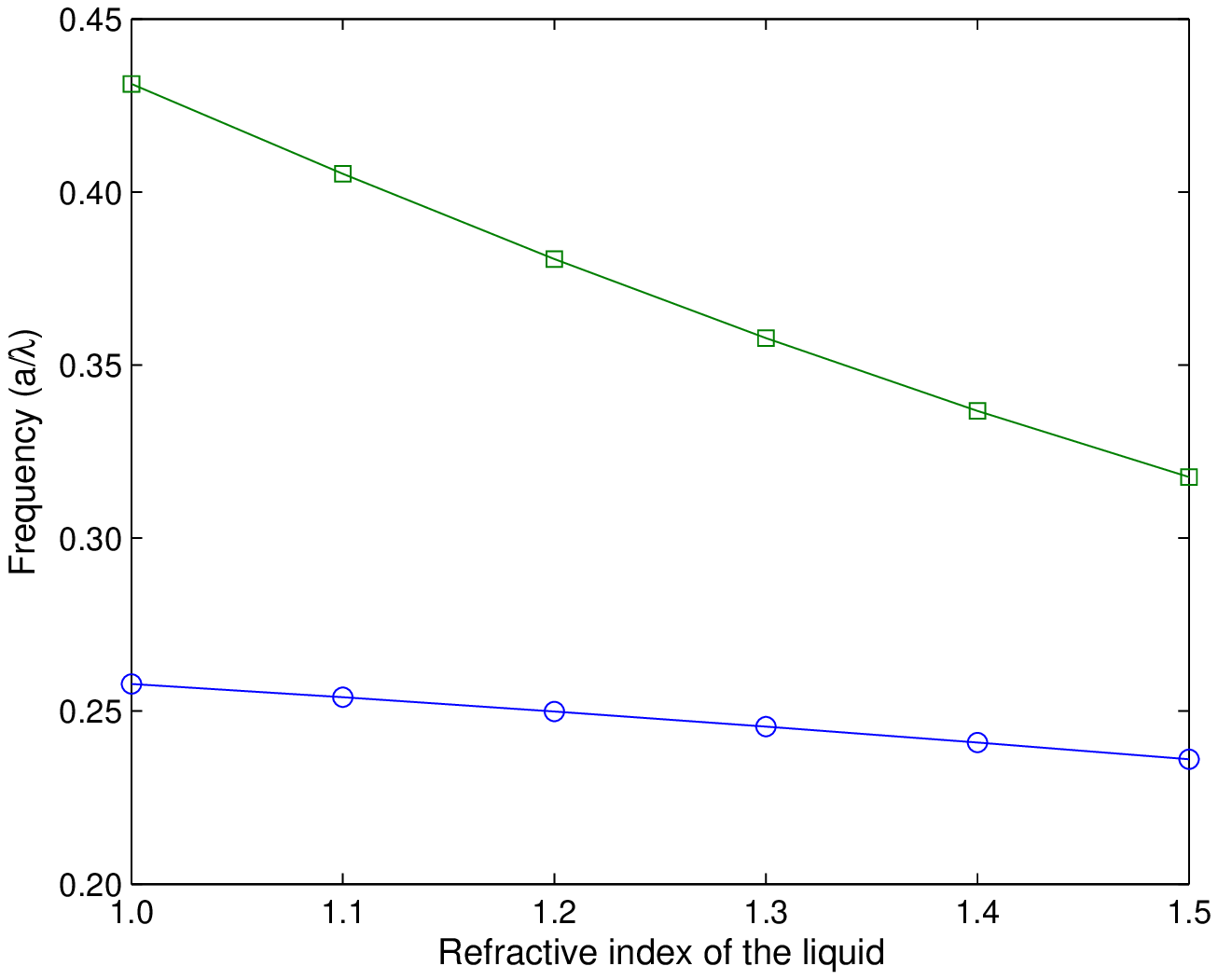,width=1\columnwidth,clip,angle=0}
\end{center}
\caption{Band-gap edges as a function of the refractive index for
the filled liquid.} \label{BandEdge-Squ}
\end{figure}

Let us next consider a square PhC with dielectric rods in air, as
shown in the inset of Fig.~\ref{Trans-Squ}. The permittivity of the
rods is $\varepsilon=10.5$, and the radius of the rods is r=0.2a.
Transmission spectra are shown in Fig.~\ref{Trans-Squ}, where the
background of the PhC is filled by different liquids. The band-gap
edge as a function of the refractive index of the liquid is shown in
Fig.~\ref{BandEdge-Squ}. Similar to the result shown in
Fig.~\ref{BandEdge-Tri}, the low-frequency band-gap edge hardly
changes as the refractive index of the liquid increases, while the
high-frequency band-gap edge is strongly dependent on the liquid.
Compared with the result for the high-frequency band-gap edge shown
in Fig.~\ref{BandEdge-Tri}, the results in Fig.~\ref{BandEdge-Squ}
illustrate a higher sensitivity. The high-frequency band-gap edge
shifts $\delta(a/\lambda)=0.025939$ when the air holes are filled by
a liquid of index $n=1.1$. For comparison we have
$\delta(a/\lambda)=0.012987$ for the structure shown in the inset of
Fig.~\ref{Trans-Tri}.

\begin{figure}[t!]
\begin{center}
\epsfig{file=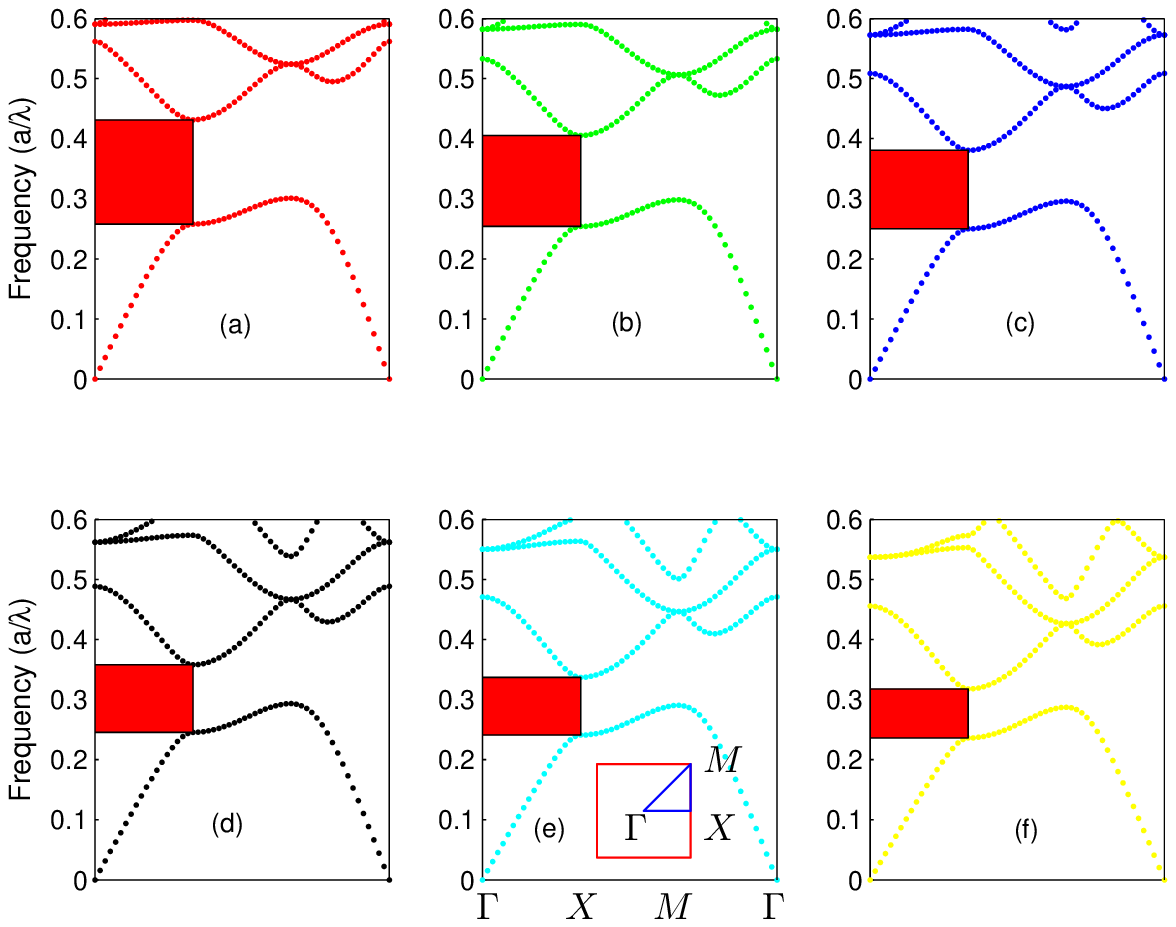,width=1\columnwidth,clip,angle=0}
\end{center}
\caption{Dispersion of the square photonic crystal shown in the
inset of Fig. 5(a), where the background is filled by liquids with
refractive indices varying from n=1.0 to 1.5 in steps of $\delta
n=0.1$. The red regions denote the bandgap regions along $\Gamma
X$ direction. } \label{Band-Squ}
\end{figure}

Band structures are shown in Fig.~\ref{Band-Squ}, where red regions
represent the band gap for $\Gamma X$ direction. From
Fig.~\ref{Trans-Squ} and Fig.~\ref{Band-Squ}, we find that the
position of the gap in the transmission spectra agree very well with
the position of the gaps for the PhC along the $\Gamma M$ direction.
When increasing the refractive index of the liquid [going from panel
(a) toward panel (f)], the high-frequency band-gap edge is
significantly downward shifted, while the low-frequency band-gap
edge slightly decreases. Again we consider a commercial silicone
fluid with a calibrated refractive-index accuracy of $\delta
n=0.002$. For the working wavelength around $1.55\,{\rm \mu m}$
($a=450\,{\rm nm}$), the mode-gap edge shifts up to 1.60~nm for
$\delta n=0.002$. The proposed biosensor relies strongly on the
dispersion of the PhC band-edge mode and the presence of a band gap.
To further improve the sensitivity, we optimize the PhC structure by
varying the radius of the rods. By a careful design of the structure
shown in the inset of Fig.~\ref{Trans-Squ}, we have been able to
improve the design further. For the working wavelength around
$1.55\,{\rm \mu m}$ ($a=450\,{\rm nm}$), the band-gap edge shifts
about 1.8~nm for $\delta n=0.002$, when $r$ is tuned to $0.1 a$.
Compared to the biosensor we proposed before~\cite{Xiao:2006}, this
structure not only shows much better sensitivity but it also seems
relatively easy to realize experimentally, since the design involves
no cavities or waveguide structures. Finally, this device, with a
size of $5\,{\rm \mu m} \times 5\,{\rm \mu m}$, is sufficiently
compact for most applications.

\section{Conclusions} To conclude, we have theoretically studied
optofluidic biosensors based on highly dispersive 2D photonic
crystal. Our study shows that these structures are strongly
sensitive to the refractive index of the liquid, which is used to
tune dispersion of photonic crystal. For the working wavelength
around $1.55{\rm \mu m}$, we predict shifts in the band-gap edge up
to 1.8~nm for $\delta n=0.002$. Although our study is based on 2D
photonic crystals, it can be extended to the case of a 2D photonic
crystal slab. For a 2D photonic crystal slab, the field will
attenuate due to out-of-plane loss, but the shift of the
high-frequency mode-gap edge at X/M point (when tuning by liquid) is
unaffected by the out-of-plane radiation. The high sensitivity makes
such devices interesting for biochemical sensing applications.

\section*{Acknowledgments}

This work is financially supported by the \emph{Danish Council for
Strategic Research} through the \emph{Strategic Program for Young
Researchers} (grant no: 2117-05-0037).

\newpage

\newpage

\begin{thebibliography}{10}
\newcommand{\enquote}[1]{``#1''}

\bibitem{John:1987}
S.~John, \enquote{Strong localization of photons in certain
disordered
  dielectric superlattices}, Phys. Rev. Lett. \textbf{58} 2486 -- 2489 (1987).

\bibitem{Yab:1987}
E.~Yablonovitch, \enquote{Inhibited spontaneous emission in solid
state physics
  and electronics}, Phys. Rev. Lett. \textbf{58} 2059--2062 (1987).

\bibitem{Joannopoulos:1995}
J.~D. Joannopoulos, R.~D. Meade, and J.~N. Winn, \emph{Photonic
crystals:
  molding the flow of light} (Princeton University Press, Princeton, 1995).

\bibitem{Lin:1996}
S.~Y. Lin, V.~M. Hietala, L.~Wang, and E.~D. Jones,
\enquote{Highly dispersive
  photonic band-gap prism}, Opt Lett \textbf{21} 1771--1773 (1996).

\bibitem{Kosaka:1998}
H.~Kosaka, T.~Kawashima, A.~Tomita, M.~Notomi, T.~Tamamura,
T.~Sato, and
  S.~Kawakami, \enquote{Superprism phenomena in photonic crystals}, Phys. Rev.
  B \textbf{58} R10096--R10099 (1998).

\bibitem{Psaltis:2006}
D.~Psaltis, S.~R. Quake, and C.~H. Yang, \enquote{Developing
optofluidic
  technology through the fusion of microfluidics and optics}, Nature
  \textbf{442} 381 -- 386 (2006).

\bibitem{Domachuk:2004}
P.~Domachuk, H.~C. Nguyen, B.~J. Eggleton, M.~Straub, and M.~Gu,
  \enquote{Microfluidic tunable photonic band-gap device}, Appl. Phys. Lett.
  \textbf{84} 1838 -- 1840 (2004).

\bibitem{Erickson:2006}
D.~Erickson, T.~Rockwood, T.~Emery, A.~Scherer, and D.~Psaltis,
  \enquote{Nanofluidic tuning of photonic crystal circuits}, Opt. Lett.
  \textbf{31} 59 -- 61 (2006).

\bibitem{Galas:2005}
J.~C. Galas, J.~Torres, M.~Belotti, Q.~Kou, and Y.~Chen,
\enquote{Microfluidic
  tunable dye laser with integrated mixer and ring resonator}, Appl. Phys.
  Lett. \textbf{86} 264101 (2005).

\bibitem{Grillet:2004}
C.~Grillet, P.~Domachuk, V.~Ta'eed, E.~Magi, J.~A. Bolger, B.~J.
Eggleton,
  L.~E. Rodd, and J.~Cooper-White, \enquote{Compact tunable microfluidic
  interferometer}, Opt. Express \textbf{12} 5440 -- 5447 (2004).

\bibitem{Kurt:2005}
H.~Kurt and D.~S. Citrin, \enquote{Coupled-resonator optical
waveguides for
  biochemical sensing of nanoliter volumes of analyte in the terahertz region},
  Appl. Phys. Lett. \textbf{87} 241119 (2005).

\bibitem{GersborgHansen:2005}
M.~Gersborg-Hansen, S.~Balslev, N.~A. Mortensen, and
A.~Kristensen, \enquote{A
  coupled cavity micro-fluidic dye ring laser}, Microelectron. Eng.
  \textbf{78-79} 185 -- 189 (2005).

\bibitem{Li:2006}
Z.~Y. Li, Z.~Y. Zhang, T.~Emery, A.~Scherer, and D.~Psaltis,
\enquote{Single
  mode optofluidic distributed feedback dye laser}, Opt. Express \textbf{14}
  696 -- 701 (2006).

\bibitem{GersborgHansen:2006}
M.~Gersborg-Hansen and A.~Kristensen, \enquote{Optofluidic third
order
  distributed feedback dye laser}, Appl. Phys. Lett. \textbf{89} 103518 (2006).

\bibitem{Chow:2004}
E.~Chow, A.~Grot, L.~W. Mirkarimi, M.~Sigalas, and G.~Girolami,
  \enquote{Ultracompact biochemical sensor built with two-dimensional photonic
  crystal microcavity}, Opt. Lett. \textbf{29} 1093 -- 1095 (2004).

\bibitem{Xiao:2006}
S.~Xiao and N.~A. Mortensen, \enquote{Highly sensitive optofluidic
biosensors
  based on dispersive photonic crystal waveguides}, preprint.

\bibitem{Xiao:2004}
S.~Xiao, M.~Qiu, Z.~Ruan, and S.~He, \enquote{Influence of the
surface
  termination to the point imaging by a photonic crystal slab with negative
  refraction}, Appl. Phys. Lett. \textbf{85} 4269--4271 (2004).

\bibitem{Ruan:2005}
Z.~Ruan, M.~Qiu, S.~Xiao, S.~He, and L.~Thylen, \enquote{Coupling
between plane
  waves and \uppercase{B}loch waves in photonic crystal with negative
  refraction}, Phys. Rev. B \textbf{71} 045111 (2005).

\bibitem{TafloveFDTD}
A.~Taflove, \emph{Computational Electrodynamics: The
Finite-Difference
  Time-Domain Method}, 2 edn. (Artech House INC, Norwood, 2000).

\bibitem{Berenger:1994}
J.~P. Berenger, \enquote{A perfectly matched layer for the
absorption of
  electromagnetic waves}, J. Comput. Phys. \textbf{114} 185--200 (1994).

\bibitem{Johnson:2001}
S.~G. Johnson and J.~D. Joannopoulos, \enquote{Block-iterative
frequency-domain
  methods for \uppercase{M}axwell's equations in a planewave basis}, Opt.
  Express \textbf{8} 173 -- 190 (2001).

\bibitem{Okamoto:2005}
K. Okamoto, M. Sugita, Y. Nagotomo, J. Yamamichi, T. Yamazaki, and
M. Uchiba, \enquote{Photonic crystal sensor with micro flow
channels}, International Symposium on Photonic and Electromagnetic
Crystal Structures (PECS-VI), June 19-24, Crete, Greece, 2005.

\end{thebibliography}
\end{document}